# Observation of anti-damping spin-orbit torques generated by in-plane and out-of-plane spin polarizations in MnPd$_3$

Mahendra DC[1]*, Ding-Fu Shao[2], Vincent D.-H. Hou[3], P. Quarterman[4], Ali Habiboglu[5], Brooks Venuti[6], Masashi Miura[1,7], Brian Kirby[4], Arturas Vailionis[8,9], Chong Bi[10], Xiang Li[1,10], Fen Xue[1], Yen-Lin Huang[3], Yong Deng[1], Shy-Jay Lin[3], Wilman Tsai[1], Serena Eley[6], Weigang Wang[5], Julie A. Borchers[4], Evgeny Y. Tsymbal[2], and Shan X. Wang[1,10]*

[1]Department of Materials Science and Engineering, Stanford University, Stanford, California 94305, USA

[2]Department of Physics and Astronomy & Nebraska Center for Materials and Nanoscience, University of Nebraska, Lincoln, NE 68588-0299

[3]Taiwan Semiconductor Manufacturing Company, Hsinchu, Taiwan

[4]NIST Center for Neutron Research, National Institute of Standards and Technology, 100 Bureau Dr., Gaithersburg, Maryland 20899, USA

[5]Department of Physics, University of Arizona, Tucson, Arizona 85721, USA

[6]Department of Physics, Colorado School of Mines, Golden, Colorado 80401

[7]Graduate School of Science and Technology, Seikei University, 3-3-1 Kichijoji-kitamachi, Musashinoshi, Tokyo, 180-8633, Japan

[8]Stanford Nano Shared Facilities, Stanford University, Stanford, CA 94305, USA

[9]Department of Physics, Kaunas University of Technology, LT-51368 Kaunas, Lithuania

[10]Department of Electrical Engineering, Stanford University, Stanford, California 94305, USA

**High spin-orbit torques (SOTs) generated by topological materials and heavy metals interfaced with a ferromagnetic layer show promise for next generation magnetic memory and logic devices. SOTs generated from the in-plane spin polarization along *y*-axis originated by the spin Hall and Edelstein effects can switch magnetization collinear with the spin polarization in the absence of external magnetic fields. However, an external magnetic field is required to switch the magnetization along *x* and *z*-axes via SOT generated by *y*-spin polarization. Here, we present that the above limitation can be circumvented by unconventional SOT in magnetron-sputtered thin film MnPd$_3$. In**

**addition to the conventional in-plane anti-damping-like torque due to the *y*-spin polarization, out-of-plane and in-plane anti-damping-like torques originating from *z*-spin and *x*-spin polarizations, respectively have been observed at room temperature. The spin torque efficiency ($\theta_y$) corresponding to the y-spin polarization from MnPd$_3$ thin films grown on thermally oxidized silicon substrate and post annealed at 400 °C is 0.34 - 0.44 while the spin conductivity ($\sigma_{zx}^{y}$) is ~ 5.70 – 7.30× 10$^5$ $\hbar/2e$ Ω$^{-1}$m$^{-1}$. Remarkably, we have demonstrated complete external magnetic field-free switching of perpendicular Co layer via unconventional out-of-plane anti-damping-like torque from *z*-spin polarization. Based on the density functional theory calculations, we determine that the observed *x*- and *z*- spin polarizations with the in-plane charge current are due to the low symmetry of the (114) oriented MnPd$_3$ thin films. Taken together, the new material reported here provides a path to realize a practical spin channel in ultrafast magnetic memory and logic devices.**

*Corresponding authors: mdc2019@stanford.edu and sxwang@stanford.edu

Efficient control of magnetization at ultra-high speed has been of prime interest to the spintronics community[1]. Spin-orbit torque (SOT) has provided efficient and ultrafast control of magnetization in magnetoresistive random access memory (MRAM) and logic devices[2,3]. SOT has been observed in magnetic semiconductor[4] and heavy metals[5–7], topological insulators[8–12], antiferromagnets[13–17], and semimetals[18–20] interfaced with a ferromagnetic layers. The charge current injected into the non-magnetic layer (spin channel) along the *x*-direction generates a spin current along *z*-direction with spin polarization ($\hat{\sigma}$) pointing along *y*-direction ($\hat{\sigma}_y$) in heavy metals due to the bulk spin Hall effect. In the case of topological insulators and Weyl semimetals non-equilibrium spin-density is accumulated at the interface due to the time reversal symmetry protected spin momentum locking[21]. SOTs exerted on the ferromagnet with in-plane magnetic anisotropy (IMA) are in-plane anti-damping-like ($\boldsymbol{\tau}_{ADL,y} \propto \hat{m} \times (\hat{\sigma}_y \times \hat{m})$) and out-of-plane field-like ($\boldsymbol{\tau}_{FL} \propto \hat{\sigma}_y \times \hat{m}$), where $\hat{m}$ is magnetization unit vector. Spin current with $\hat{\sigma}$ along *z*-direction ($\hat{\sigma}_z$) has been observed in transition-metal dichalcogenides[18,20,22] and, ferromagnets interfaced with light metals[23] with the charge current flow along *x*-direction. SOT due to $\hat{\sigma}_z$ exerts torque along the out-of-plane direction ($\boldsymbol{\tau}_{ADL,z} \propto \hat{m} \times (\hat{\sigma}_z \times \hat{m})$), which will enable external field free and low power switching of the out-of-plane magnetization[24]. Recently, $\hat{\sigma}$ along the *x*-direction ($\hat{\sigma}_x$) has been reported in the uncompensated antiferromagnet Mn$_3$GaN in addition to $\hat{\sigma}_y$ and $\hat{\sigma}_z$ due to the low magnetic symmetry[17]. SOT due to $\hat{\sigma}_x$ exerts torque along in-plane direction ($\boldsymbol{\tau}_{ADL,x} \propto \hat{m} \times (\hat{\sigma}_x \times \hat{m})$), which will deterministically switch the magnetization along the *x*-direction in the absence of external magnetic field. The figure of merit of charge to spin conversion is known as the spin torque efficiency $\theta_k \propto \frac{\sigma_{ij}^k}{\sigma_{xx}}$, where $\sigma_{ij}^k$ (*i*, *j*, and *k* refers to the spin current flow, charge current flow, and spin polarization directions, respectively) and $\sigma_{xx}$ are spin and charge conductivities, respectively. $\theta_k$ needs to be high for efficient control of the magnetization. Furthermore, to avoid current shunting through a conducting ferromagnetic layer high $\sigma_{ij}^k$ is also required[25]. Another important requirement for the integration of a spin channel into semiconductor IC technology is the tolerance of SOT materials to thermal annealing at 400 °C. However, there has not been a practical spin channel which can handle post annealing at that temperature, and also possesses high $\tau_{ADL,y}$ along with $\tau_{ADL,x}$ and $\tau_{ADL,z}$, enabling deterministic switching of in-plane magnetization along *y*, in-plane magnetization along *x*, and out-of-plane magnetization, respectively, without the need of applying an external magnetic field.

To achieve high density magnetic memory and logic devices, perpendicular magnetic anisotropy (PMA) is desired[26]. PMA switching via SOT from heavy metals[5–7] and topological insulators[11,12,27,28] has been reported at the room temperature in the presence of an external magnetic field. Partial PMA switching has been observed on PtMn/[Co/Ni]$_x$[13], IrMn/CoFeB[14], and PtMn/CoFeB/Gd/CoFeB[16] stack structures, in the absence of external magnetic field, but with the help of exchange bias. Stray fields from an in-plane magnetic layer present above or below the spin channel can facilitate external magnetic field-free PMA switching, but current shunting and magnetic interference between different magnetic layers pose severe design constraints in this approach[29,30]. Combination of SOT and spin transfer torque (STT) can also switch PMA in the absence of external magnetic field, but STT could lead to reduced endurance of magnetic tunnel barrier and slower magnetization switching[31]. Fast switching of the magnetization with lower critical current densities ($J_{sw}$) can be achieved when the charge current flow and magnetization are collinear, however, this geometry still requires an external magnetic field to achieve magnetization switching[7]. Here, we present SOT from sputtered MnPd$_3$ thin films post annealed at 400 °C that can generate a spin current with $\hat{\sigma}$ along all three axes due to the charge current flow along *x*-direction, which represents a major advance over the literature and overcomes significant limitations of the existing SOT materials.

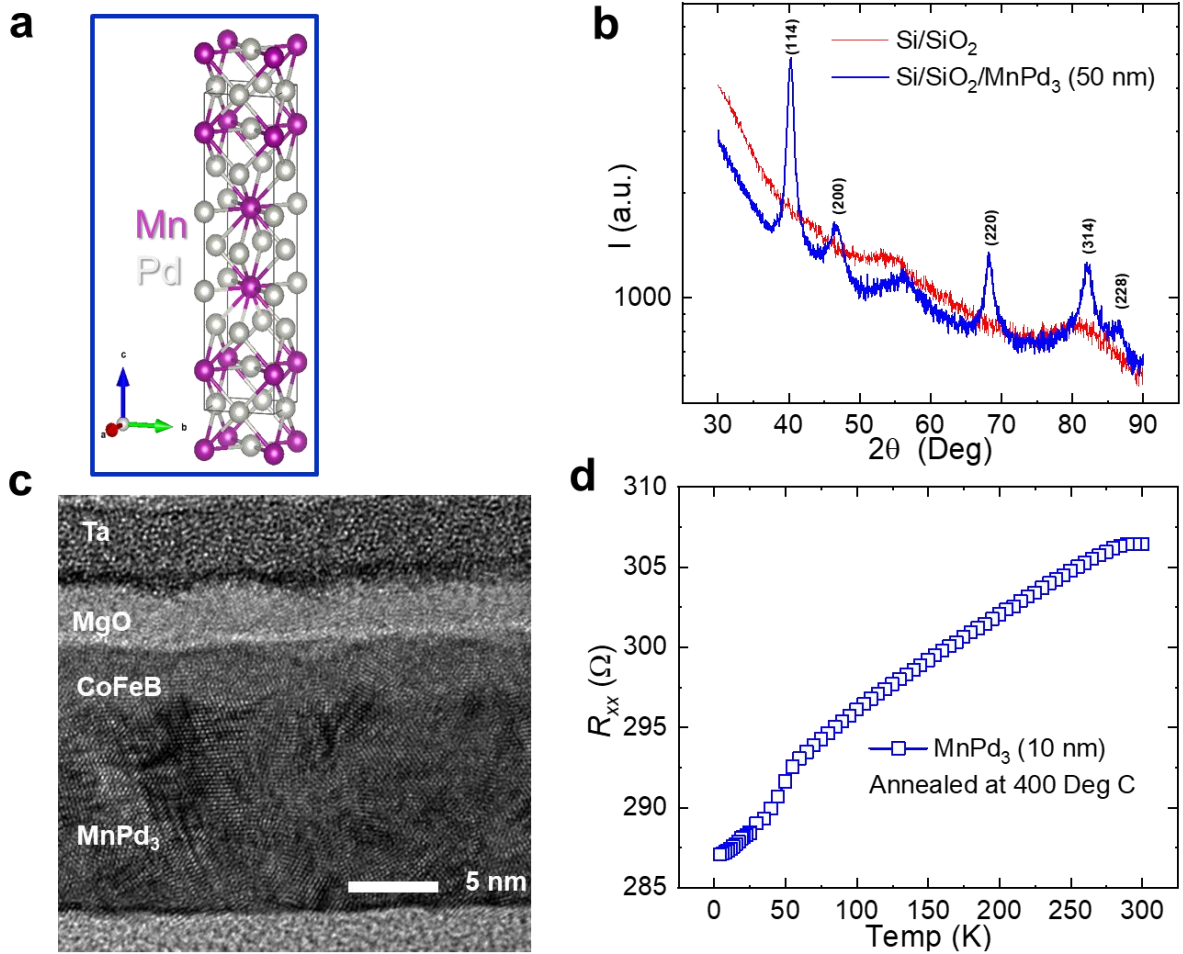

**Fig. 1. Characterization of MnPd₃ thin film**: **a,** Schematic diagram of D023 MnPd$_3$ unit cell. **b,** XRD of Si/SiO$_2$/MnPd$_3$ (50 nm) film, post-annealed at 400 ºC for 30 min. **c,** Cross-section TEM image of Si/SiO$_2$/MnPd$_3$ (10 nm)/CoFeB (5 nm)/MgO (2 nm)/Ta (2 nm) sample. **d,** Four terminal resistance as a function of temperature of Si/SiO$_2$/MnPd$_3$ (10 nm) sample.

The MnPd$_3$ thin films were magnetron sputtered at room temperature on 300 nm thick thermally oxidized silicon substrates. The thin films with the stack structure Si/SiO$_2$/MnPd$_3$ (*t* nm)/CoFeB (5 nm)/MgO (2 nm)/Ta (2 nm) with IMA were prepared for the SOT characterization with *t* = 4, 6, 8, 10, 12, 16, 20, and 24 nm, respectively. Unless otherwise stated, these films will be labeled MP4 - MP24, in which the number denotes the MnPd$_3$ thickness. All samples were post annealed at 400 °C for 30 minutes. Using Rutherford backscattering, the atomic composition of Mn and Pd in MnPd$_3$ film is 28% and 72% (data not shown), respectively.

Fig. 1a shows the unit cell of MnPd$_3$. We performed grazing incidence $\theta$ - $2\theta$ X-ray diffraction (XRD) measurements on a Si/SiO$_2$/MnPd$_3$ (50 nm) sample, as shown in Fig. 1b. For the grazing incidence XRD measurement $\Phi$ and $\Omega$ were fixed at 20° and 3.5°, respectively. From the intensity of the peaks and pole figures (Methods and Extended Data Fig. 1), its notable that MnPd$_3$ film has a strong (114) texture. The lattice parameters are estimated to be a = 3.89 Å, b = 3.88 Å, and c = 15.42 Å indexed by using ref. ([32]). The cross-section transmission electron microscopy (TEM) bright image of the MP10 sample is presented in Fig. 1c. The high angle annular dark field (HAADF) image (data not shown) and the bright image both show that the MnPd$_3$ layer grown on thermally oxidized silicon is polycrystalline. The CoFeB and MgO layers are also polycrystalline. The electric and magnetotransport measurements were performed on Si/SiO$_2$/MnPd$_3$ (10 nm)/MgO (2 nm)/Ta (2 nm) heterostructure, it will be labelled as MnPd (10 nm) sample (Methods and Extended Data Fig. 2). The resistivity shows metallic behavior coinciding with the possible transition from a paramagnetic to an antiferromagnetic state below 50 K, as shown in Fig. 1d. The ordinary Hall resistance as a function of the external magnetic field is non-linear at small fields. From the high-field linear region, we estimate a carrier concentration of 4.4 × 10$^{22}$/cm$^3$. At room temperature the values of anisotropic magnetoresistance (AMR) and planar Hall resistance (PHR) are estimated to be 0.012% and 20 mΩ in MnPd (10 nm) sample, respectively. The Néel temperature of MnPd (10 nm) sample is approximately 37 K inferred from using temperature-dependent magnetometry (Methods and Extended Data Fig. 3). Polarized neutron reflectometry (PNR) measurements show weak ferromagnetism, persisting upto room temperature, in MnPd$_3$ films possibly originating from uncompensated Mn moments, Mn clusters, or local ferromagnetic Mn-based compound formation. At room temperature the ferromagnetic component of the magnetization in MnPd$_3$ is determined to be 9 ± 1.9 kA/m using polarized neutron reflection (PNR) whereas at 6 K it is ~ 43 ± 3.4 kA/m (Methods and Extended Data Fig. 4).

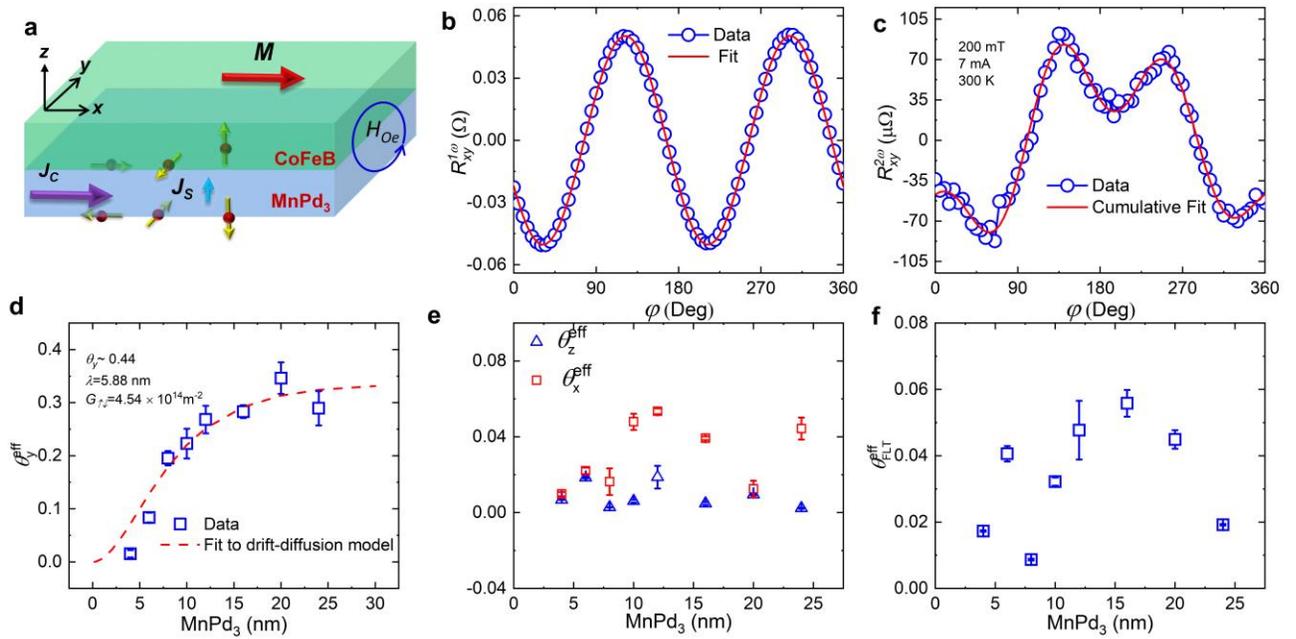

**Fig. 2. SOT characterization using Second Harmonic Hall (SHH) technique on Si/SiO$_2$/MnPd$_3$ (x nm)/CoFeB (5 nm)/MgO (2 nm)/Ta (2 nm): a,** Schematic diagram showing in-plane charge current generated spin current with spin polarizations along three axes. The red

spheres represent electrons and yellow arrows represent spin magnetic moment, respectively. **b,** and **c,** $R_{xy}^{1\omega}$ and $R_{xy}^{2\omega}$, respectively, as a function of in-plane magnetic field rotation at a fixed amplitude of 200 mT in MP12 sample. **d**, **e,** and **f,** Effective spin torque efficiency due to the in-plane anti-damping-like, in-plane and out-of-plane anti-damping-like and field-like torques, respectively, as a function of $MnPd_3$ film thickness. The Hall bar device used for this measurement was 10 μm wide and 130 μm long, respectively.

We performed SOT measurements using SHH technique on MP4-MP24 samples, control MP (10 nm) and Si/SiO$_2$/ CoFeB (5 nm)/MgO (2 nm)/Ta (2 nm) samples, and a reference Si/SiO$_2$/Pt (10 nm)/CoFeB (5 nm)/MgO (2 nm)/Ta (2 nm) ( labelled as Pt10 sample) sample[33]. The MP4-MP24 samples and reference sample were patterned into Hall bars with length 130 and width 10 μm, respectively. The details of the SHH are presented in Methods and Extended Data Fig. 5. The a.c. current injected into the Hall bar induces effective spin-orbit fields, which oscillate the magnetization around its equilibrium position, and as a result, a SHH voltage is induced. In the SHH measurement, the sample is rotated in the *x-y* plane under constant static magnetic field to keep the magnetization in a single domain state. The spin-current with the spin magnetic moment pointing along the negative *y*-axis and positive *x*-and *z*-axes get accumulated between MnPd$_3$ and CoFeB layers as shown in Fig. 2a. Thus, accumulated spin-currents exert $\boldsymbol{\tau}_{ADL,x}$, $\boldsymbol{\tau}_{ADL,y}$ and $\boldsymbol{\tau}_{ADL,z}$, $\boldsymbol{\tau}_{FL}$ on the CoFeB layer along the in-plane and out-of-plane directions, respectively. In addition, the Oersted field ($H_{Oe}$) generated due to the a.c. current flow in the MnPd$_3$ layer exerts an Oersted torque ($\boldsymbol{\tau}_{Oe}$) on the CoFeB layer. Fig. 2b shows $R_{xy}^{1\omega}$ as a function of the in-plane magnetic field angle ($\varphi$). $R_{xy}^{1\omega}$ fits perfectly to the $sin2\varphi$, indicating that the out-of-plane field projection due to the imperfect sample mounting is absent. $R_{xy}^{2\omega}$ as a function of magnetic field angle is presented in Fig. 3c. Since the torques have different dependencies with $\varphi$, we can extract $R_{xy}^{2\omega}$ due to different types of torques using SHH. In MP10 sample, the extracted spin-orbit fields associated with the $\tau_{ADL,x}$, $\tau_{ADL,y}$, $\tau_{ADL,z}$, and $\tau_{FL}$ are (0.02 ± 0.002), (0.132 ± 0.002), (0.0036 ± 0.0003), and (0.019 ± 0.00) mT per $10^6$ A/cm$^2$, respectively. These values of the $\tau_{ADL,y}$ and $\tau_{FL}$ are comparable or better than the previous reports on heavy metals/ferromagnet[33], TIs/ferromagnet[11,12], and Weyl semimetal/ferromagnets[19]. In Fig. 2d $\theta_y^{eff}$ as a function of MnPd$_3$ film thickness is presented, which shows heavy metal like behavior. The simple drift-diffusion model (Eqn. 1) can be utilized to extract bulk spin-torque efficiency figure of merit ($\theta_y(t \approx \infty)$) and spin-diffusion length ($\lambda$),

$$\theta_y^{eff} = \theta_y(t \approx \infty)(1 - \text{sech}\left(\frac{t}{\lambda}\right)) \quad (1)$$

where $t$ is MnPd$_3$ film thickness. This drift-diffusion model considers that the spin current generated by the bulk of thin films is completely absorbed by the ferromagnetic layer without any dissipation at the interface and back-flow of the spin current. $\theta_y(t \approx \infty)$ and $\lambda$ obtained by ordinary drift-diffusion model are 0.34 and 6.30 nm, respectively. Now by considering spin-back flow the drift-diffusion model can be modified into[34–36]:

$$\theta_y^{eff} = \theta_y(t \approx \infty)(1 - \text{sech}(\frac{t}{\lambda}))[\frac{1+\tanh(\left(\frac{t}{2\lambda}\right))}{2\rho\lambda G_{\uparrow\downarrow}}]^{-1} \quad (2)$$

where $\rho$ is bulk resistivity, $G_{\uparrow\downarrow}$ is spin-mixing conductivity, respectively. The red line in Fig. 2d is a fit to Eqn. (2) with $\theta_y(t \approx \infty)$ and $\lambda$ as independent fitting parameters. $G_{\uparrow\downarrow}$ values of MP4 and MP24 samples are estimated to be 4.54 × 10$^{14}$ m$^{-2}$ and 3.70 × 10$^{15}$ m$^{-2}$, respectively. The extracted values of $\theta_y(t \approx \infty)$ and $\lambda$ are 0.44 and 5.88 nm, respectively for $G_{\uparrow\downarrow}$ value of 4.54 ×

$10^{14}$ m$^{-2}$. $\rho$ value used for the fitting was 60 μΩcm obtained by measuring four terminal resistance of MnPd$_3$ (20 nm) sample. The extracted values of $\theta_y(t \approx \infty)$ and $\lambda$ are 0.36 and 6.30 nm, respectively for $G_{\uparrow\downarrow}$ value of $3.70 \times 10^{15}$ m$^{-2}$. The values of the $\sigma_{zx}^y$ corresponding to $\theta_y$ (0.34 - 0.44) are estimated to be $(5.67 - 7.33) \times 10^5$ $\hbar/2e$ Ω$^{-1}$m$^{-1}$. These values of $\sigma_{zx}^y$ are among the largest for the reported values of the antiferromagnets[15,37], heavy metals[6,38,39], topological insulators[8,11], and Weyl semimetals[19]. $R_{xy}^{2\omega}$ in the control samples does not show any field or $\varphi$ dependence indicating that the SOTs in the MP samples originates from the MnPd$_3$ layer (Methods and Extended Data Fig. 5 and Fig. 6). The values for the spin-orbit fields associated with $\tau_{ADL,y}$ and $\tau_{FL}$ of the reference Pt10 sample are estimated to be $(0.040 \pm 0.003)$ and $(0.012 \pm 0.001)$ mT per $10^6$ A/cm$^2$, respectively (Methods and Extended Data Fig. 7). The estimated value of $\theta_y^{eff}$ and $\theta_{FL}^{eff}$ are $(0.07 \pm 0.01)$ and $(0.02 \pm 0.002)$, respectively in agreement with the previous reports[6,33,40]. After post annealing at 400 °C, the reference Pt10 sample does not show a SHH signal, suggesting that its SOT did not withstand such annealing.

As presented in Fig. 2e, $\theta_z^{eff}$ and $\theta_x^{eff}$ do not depend on the MnPd$_3$ film thickness as $\theta_y^{eff}$ does. $\theta_{FL}^{eff}$ shown in Fig. 2d also does not show any specific MnPd$_3$ film thickness dependence. The $\sigma_{zx}^{z,eff}$ value in MP12 sample is as large as ~ $0.14 \times 10^5$ $\hbar/2e$ Ω$^{-1}$m$^{-1}$. $\sigma_{zx}^{z,eff}$ in MP samples is comparable or better than recent reports on WTe$_2$/Py[18] and Mn$_3$GaN/Py[17]. $\sigma_{zx}^{x,eff}$ in MP12 sample is $0.77 \times 10^5$ $\hbar/2e$ Ω$^{-1}$m$^{-1}$. These values of $\theta_k$ and $\sigma_{ij}^k$ are largest among the reported values as listed in Table 1. The difference in the thickness dependence of the $\theta_{x,y,z}^{eff}$ indicate that their origins are also different.

We also performed ST-FMR measurement on MP10 sample as a confirmation of the observed SOT with SHH technique. The details of ST-FMR are presented in Methods and Extended Data Fig. 8. $\theta_y^{eff}$ estimated by using ST-FMR of MP10 sample is (0.21 ± 0.01) whereas it is (0.22 ± 0.03) estimated using SHH. We also prepared reference samples Si/SiO$_2$/Pt (6 and 10 nm)/CoFeB (5 nm)/MgO (2 nm)/Ta (2 nm) (labelled as Pt6 and Pt10 samples) and Si/SiO$_2$/W (6 nm)/CoFeB (2 nm)/MgO (2 nm)/Ta (2 nm) (labelled as W sample) for the ST-FMR measurements. $\theta_y^{eff}$ of as deposited Pt samples is 0.06 ± 0.01 and 0.08 ± 0.01, respectively (Methods and Extended Data Fig. 9). However, after post annealing at 400 °C the reference Pt samples do not show a ST-FMR signal, suggesting that SOT did not withstand the annealing process. $\theta_y^{eff}$ of as deposited W sample is determined to be -0.43 ± 0.03 at 6 GHz excitation frequency. The estimated $\sigma_{zx}^{y,eff}$ is -1.43 × 10$^5$ $\hbar/2e$ Ω$^{-1}$m$^{-1}$. This value of $\theta_y^{eff}$ of the as deposited W sample is comparable to the previous report[41]. $\theta_y^{eff}$ of the W sample post annealed at 400 °C for 30 minutes is estimated to be -0.011 and the corresponding $\sigma_y^{eff}$ is -0.13 × 10$^5$ $\hbar/2e$ Ω$^{-1}$m$^{-1}$ (Methods and Extended Data Fig. 9).

| Materials | $\rho$ ($\mu\Omega$cm) | $\theta_x^{eff}$ | $\theta_y$ | $\theta_z^{eff}$ | $\theta_{FL}^{eff}$ | $\sigma_{zx}^y$($\hbar/2e$ $10^5$ $\Omega^{-1}$m$^{-1}$) | $J_{sw}$ (MA/cm$^2$) | Switched magnetic anisotropy | $H_x$ (mT) | Post annealing temp (°C) |
|---|---|---|---|---|---|---|---|---|---|---|
| This work | 60 - 95 | 0.053 | 0.34-0.44 | 0.018 | 0.06 | 5.7-7.3 | 11-24.7 | IMA-PMA | Zero | 400 |
| Pt[40] | 40 | - | 0.14 | - | 0.08 | 3.5 | 50 | PMA | 70 | As dep. |
| W[38] | 188 | - | -0.22 | - | -0.022 | -1.2 | 11[42] | PMA | 20 | 200[42]-300[38] |
| MnGaN[17] | 225 | -0.013 | 0.025 | 0.018 | -0.15 | 0.11 | - | - | - | As dep. |
| WTe$_2$[18] | 380 | - | 0.03 | 0.013 | 0.034 | 0.08 | - | - | - | As dep. |

**Table 1: Summary of $\theta_k$, $\sigma_{ij}^k$, and $\rho$ of different spin channels post annealed at different temperatures and measured at room temperature.**

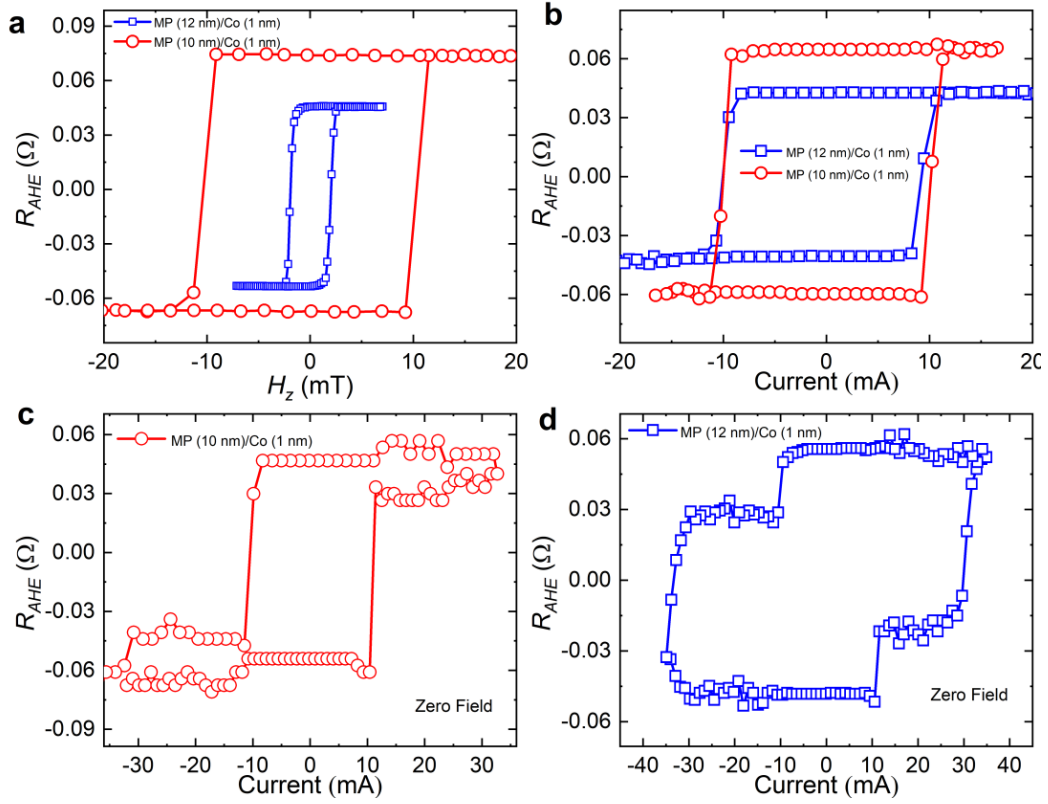

**Fig. 3. Demonstration of external magnetic field-free out-of-plane magnetization switching via *z*-spin polarization generated anti-damping spin-orbit torque:** **a,** The anomalous Hall resistance as a function of out-of-plane external field. **b,** Switching of perpendicular Co layer via SOT under the application of -8 mT and -20 mT along *x*-direction for MP (10 nm)/Co (1 nm) and MP (12 nm)/Co (1 nm) samples, respectively. **c,** and **d,** Field-free perpendicular Co layer switching via out-of-plane anti-damping-like torque generated by *z*-spin polarization. The Hall bar with a length of 130 μm and a width of 10 μm was used for the magnetization switching experiment.

In order to demonstrate external magnetic field-free PMA switching, we prepared Si/SiO$_2$/MnPd$_3$ (10 and 12 nm)/Co (1 nm)/MgO (2 nm)/Ta (2 nm) samples (will be labelled as MP10/Co1 and MP12/Co1 samples). MP10/Co1 and MP12/Co1 samples were annealed at 400 ºC for 30 minutes in vacuum and subsequently field-cooled under the application of an out-of-plane magnetic field of 0.45 T. Fig. 3a shows anomalous Hall resistance ($R_{AHE}$) as a function of out-of-plane magnetic field. The hysteretic $R_{AHE}$ loop confirms PMA is present in the Co layer. Alternatively, magnetometry was also used to confirm that PMA is present in MP10/Co1 sample (Methods and Extended Data Fig. 3). Fig. 3b shows current-induced SOT magnetization switching under the presence of negative external magnetic fields. The write d.c. current pulse width used for PMA switching is 20 ms, which is followed by a read current of 0.4 mA. The full switching of magnetization occurs in MP10/Co1 and MP12/Co1 samples at ~ 10.1 mA and 9.5 mA, respectively. Then, in the absence of external magnetic field, the d.c. current in pulses is swept from -36 mA to + 36 mA with a step size of 1.06 mA. For MP10/Co1 as shown in Fig. 3c, partial switching of the magnetization occurs at a positive current of about 10.78 mA, and complete switching occurs at ~ 32.4 mA (~37.0 MA/cm$^2$). In the subsequent reverse sweep partial switching of magnetization occurs at ~ -10.87 mA and complete switching occurs at ~ -31.63 mA. For MP12/Co1 sample as shown in Fig. 3d, partial switching of magnetization occurs at a positive current of ~ 11.35 mA. Continuously sweeping the d.c. pulses switches the remaining magnetization at ~ 30.92 mA (~24.7 MA/cm$^2$). Subsequently reverse sweeping the current pulses, partial switching of magnetization occurs at ~ - 9.60 mA and switching of remaining magnetization occurs at ~ – 33.32 mA. Since the $R_{AHE}$ values obtained by field sweep and current sweep are close, we can conclude that the full switching of PMA has been observed in both PMA samples. $J_{sw}$ values observed in our PMA samples without external magnetic field is comparable or better than the previously reported values in Pt/Co[5,6] ( ~23-100 MA/cm$^2$), Pd$_{0.25}$Pt$_{0.75}$/Co[43] (~22 MA/cm$^2$), Pt/antiferromagnet[44] with external magnetic field. The SOT switching of magnetization in our PMA samples results from the interplay of $\tau_{ADL,x}$, $\tau_{ADL,y}$,

$\tau_{ADL,z}$, and $\tau_{FL}$. In the presence of external magnetic field the $R_{AHE}$ vs $I$ loop shows similar behavior as that of positive $\theta_y^{eff}$ such as in the case of Pt/Co/AlOx[5,6]. In the absence of an external magnetic field, if there is only $\tau_{ADL,z}$, PMA switching occurs via anti-damping process, which is confirmed by numerically solving Landau-Lifshitz-Gilbert (LLG) equation (Methods and Extended Data Fig. 10b). In the presence of a relatively weaker $\tau_{ADL,z}$ and $\tau_{ADL,x}$ and strong $\tau_{ADL,y}$ the magnetization is partially switched at a lower current (~10 mA), which results in an intermediate state. The intermediate state can occur due to an insufficient external magnetic field, which is unable to completely break mirror symmetry[45]. Previously, intermediate magnetic states were observed below threshold $J_{sw}$ (Ref. [46]). The external magnetic field-free switching of PMA unambiguously demonstrates the presence of $\hat{\sigma}_z$ generated $\tau_{ADL,z}$ in the MP samples. These experimental results of PMA switching have been qualitatively reproduced by the LLG simulations (Methods and Extended Data Fig. 10c and 10d). The intermediate states observed could be utilized for neuromorphic computing[13]. In addition to field-free PMA switching, we also performed field-free magnetization switching of the in-plane CoFeB layer in MP24 sample, as detected by using unidirectional spin Hall magnetoresistance (USMR) mechanism[47,48] (Methods and Extended Data Fig. 11). $J_{sw}$ is estimated to be ~11.0 MA/cm$^2$ using the parallel resistor model.

We also performed SHH measurements on the PMA samples, as detailed in Methods and Extended Data Fig. 13. The magnitude of the SHH resistance ($R_{xy}^{2\omega}$) is not symmetric at up and down magnetizations when the field is swept along $y$-axis. If there were only $\tau_{FL}$ and $\tau_{Oe}$ present in MP samples, the magnitude and field dependence of $R_{xy}^{2\omega}$ would remain the same since the spin-orbit field associated to them is independent of the magnetization polarity as in the case of Ta/CoFeB/MgO[23]. The spin-orbit field associated with $\tau_{ADL,x}$ ($H_{ADL,x} \sim (\hat{\sigma}_x \times \hat{m})$) switches sign as the magnetization switches sign, which results in the unequal and different field dependence of $R_{xy}^{2\omega}$. This clearly shows the presence of torque generated by $\hat{\sigma}_x$ in MP samples. $\theta_x^{eff}$, $\theta_y^{eff}$, and $\theta_{FL}^{eff}$ of the MP10/Co1 sample (MP12/Co1 sample) are 0.023 ± 0.001 (0.040 ± 0.003), 0.24 ± 0.02 (0.32 ± 0.03), and 0.03 ± 0.001 (0.12 ± 0.01), respectively. These values are comparable to the SHH-measured SOT efficiencies of IMA MP10 and MP12 samples.

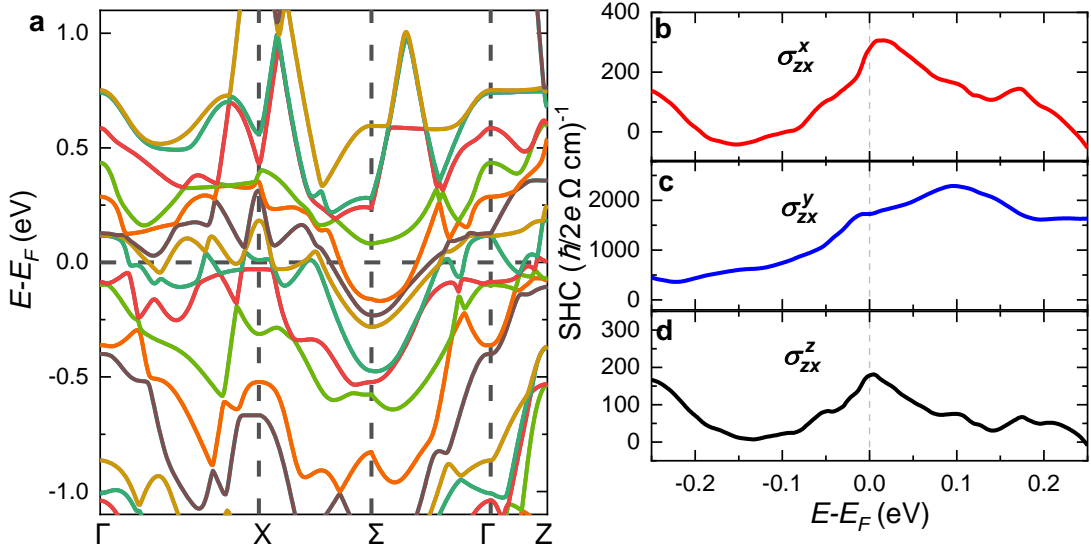

**Fig. 4: Effect of (114) texture on the spin polarization. a,** The calculated band structure of stoichiometric MnPd$_3$ at room temperature. **b**, **c**, and **d,** The calculated $\sigma_{zx}^x$, $\sigma_{zx}^y$, and $\sigma_{zx}^z$ as a function of energy for MnPd$_3$ (114) film, where the $x$ axis is oriented along the $[\bar{4}01]$ direction.

We explain the appearance of the non-vanishing $\hat{\sigma}_x$ and $\hat{\sigma}_z$ in terms the contribution from grains of different orientations in our polycrystalline MnPd$_3$ films. MnPd$_3$ has crystal space group I4/mmm [32]. If the film was monocrystalline and (001) oriented with the charge current flowing along the [100] direction (the *x*-direction), only the conventional spin Hall conductivity $\sigma_{zx}^y$ (where the spin current flows along the [001] direction (the *z*-direction) normal to the charge current and has spin polarization along the [010] direction (the *y*-direction)), would be allowed due to the (001) plane being invariant to all symmetry operations of this space group. In a polycrystalline film, however, other crystal orientations with lower symmetries contribute to the spin Hall conductivity. For example, the (114) plane corresponding to the dominant texture of our films (reflected by the strongest XRD peak in Fig. 1 (b)) is only invariant with respect to mirror symmetry $M_{[\bar{1}10]}$ and two-fold rotation $C_{[\bar{1}10]}$. This allows the appearance of unconventional components of the spin Hall conductivity tensor, such as $\sigma_{zx}^{x}$ and $\sigma_{zx}^{z}$, where the spin polarization is parallel to the direction of the spin current or the charge current. Moreover, in a polycrystalline film, the current direction itself varies with respect to the high symmetry directions of different grains, which also influences the shape of the spin Hall conductivity tensor. These aspects have been discussed in Ref. [15].

To quantitatively evaluate the contribution from this mechanism, we perform first-principles density functional theory calculations of spin Hall conductivity of bulk MnPd$_3$ assuming a room-temperature paramagnetic phase. Figure 4a shows the calculated band structure of MnPd$_3$. There are several bands crossing the Fermi level ($E_F$), indicating the metallic ground state. The small gaps between the bands near $E_F$ are favorable for the sizable spin Hall conductivity[49], which is given by[50]

$$\sigma_{ij}^{k} = \frac{e^2}{\hbar} \int \frac{d^3\vec{k}}{(2\pi)^3} \sum_n f_{n\vec{k}} \Omega_{n,ij}^{k}(\vec{k}) ,  \qquad (3)$$

$$\Omega_{n,ij}^{k}(\vec{k}) = -2Im \sum_{n' \neq n} \frac{\langle n\vec{k}|J_i^k|n'\vec{k}\rangle \langle n'\vec{k}|v_j|n\vec{k}\rangle}{\left(E_{n\vec{k}} - E_{n'\vec{k}}\right)^2} , \qquad (4),$$

where $f_{n\vec{k}}$ is the Fermi-Dirac distribution function for band $n$ and wave vector $\vec{k}$, $\Omega_{n,ij}^{k}(\vec{k})$ is the spin Berry curvature, $J_i^k = \frac{1}{2}\{v_i, s_k\}$ is the spin-current operator, $v_i$ and $s_k$ are velocity and spin operators, respectively, and $i,j,k = x,y,z$. As expected, for MnPd$_3$ textured in the (001) plane, only the conventional spin Hall conductivity ($\sigma_{zx}^y$) is non-vanishing (Table 2 in Extended Data). However, for the dominant (114) stacking texture, the unconventional spin Hall conductivities ($\sigma_{zx}^x$ and $\sigma_{zx}^z$) emerge (Table 2 in Extended Data). Figures 4b-d show the calculated spin Hall conductivities for MnPd$_3$ (114) film as a function of energy when the charge current flows along the [$\bar{4}$01] direction (the *x*-direction). We find a high conventional $\sigma_{zx}^y \sim 1744$ $(\frac{\hbar}{2e})(\Omega\text{ cm})^{-1}$ and sizable unconventional conductivities $\sigma_{zx}^x \sim 279$ $(\frac{\hbar}{2e})(\Omega\text{ cm})^{-1}$ and $\sigma_{zx}^z \sim 166$ $(\frac{\hbar}{2e})(\Omega\text{ cm})^{-1}$ at the Fermi energy. It is evident that the $\sigma_{zx}^x$ and $\sigma_{zx}^z$ values are approximately an order of magnitude smaller than $\sigma_{zx}^y$, which is consistent with our experimental observation. Similarly, other MnPd$_3$ grains with different orientations can also contribute to the unconventional spin Hall conductivity.

In summary, we studied anti-damping spin-orbit torques generated by the $\hat{\sigma}_x$, $\hat{\sigma}_y$, and $\hat{\sigma}_z$ in MnPd$_3$/ferromagnet heterostructure. At least two independent characterizations were performed to verify the presence of torques. DFT simulations confirmed the low crystal symmetry present in the (114) oriented MnPd3 thin films as the origin of the observed unconventional SOTs. We demonstrated successful growth of conductive MnPd$_3$ thin films with high $\sigma_{zx}^x$, $\sigma_{zx}^y$, and $\sigma_{zx}^z$ after post annealing at 400 °C for half an hour. Complete external magnetic field-free switching of both IMA and PMA were realized. The observed SOTs were robust against thermal treatment, and compatible with low damping constant of CoFeB even after post annealing. All of these are key factors for the integration of a practical spin current source based on MnPd$_3$ into next generation of SOT-based spintronics devices.

**References**

1. Dieny, B. *et al.* Opportunities and challenges for spintronics in the microelectronics industry. *Nat. Electron.* **3**, 446–459 (2020).

2. Grimaldi, E. *et al.* Single-shot dynamics of spin–orbit torque and spin transfer torque switching in three-terminal magnetic tunnel junctions. *Nat. Nanotechnol.* **15**, 111–117 (2020).

3. Manchon, A. *et al.* Current-induced spin-orbit torques in ferromagnetic and antiferromagnetic systems. *Rev. Mod. Phys.* **91**, 035004 (2019).

4. Chernyshov, A. *et al.* Evidence for reversible control of magnetization in a ferromagnetic material via spin-orbit magnetic field. *Nat. Phys.* **5**, 656 (2008).

5. Miron, I. M. *et al.* Perpendicular switching of a single ferromagnetic layer induced by in-plane current injection. *Nature* **476**, 189–193 (2011).

6. Liu, L., Lee, O. J., Gudmundsen, T. J., Ralph, D. C. & Buhrman, R. A. Current-induced switching of perpendicularly magnetized magnetic layers using spin torque from the spin hall effect. *Phys. Rev. Lett.* **109**, 096602 (2012).

7. Fukami, S., Anekawa, T., Zhang, C. & Ohno, H. A spin-orbit torque switching scheme with collinear magnetic easy axis and current configuration. *Nat. Nanotechnol.* **11**, 621–626 (2016).

8. Mellnik, A. R. *et al.* Spin-transfer torque generated by a topological insulator. *Nature* **511**, 449–451 (2014).

9. Fan, Y. *et al.* Magnetization switching through giant spin-orbit torque in a magnetically

doped topological insulator heterostructure. *Nat. Mater.* **13**, 699–704 (2014).

10. Kondou, K. *et al.* Fermi level dependent charge-to-spin current conversion by Dirac surface state of topological insulators. *Nat. Phys.* **12**, 1027–1032 (2016).

11. DC, M. *et al.* Room-temperature high spin-orbit torque due to quantum confinement in sputtered $Bi_xSe_{(1-x)}$ films. *Nat. Mater.* **17**, 800–807 (2018).

12. Khang, N., Khang, D., Ueda, Y. & Hai, P. N. A conductive topological insulator with large spin Hall effect for ultralow power spin-orbit torque switching. *Nat. Mater.* **17**, 808–813 (2018).

13. Fukami, S., Zhang, C., DuttaGupta, S., Kurenkov, A. & Ohno, H. Magnetization switching by spin–orbit torque in an antiferromagnet–ferromagnet bilayer system. *Nat. Mater.* **15**, 535–541 (2016).

14. Oh, Y. W. *et al.* Field-free switching of perpendicular magnetization through spin-orbit torque in antiferromagnet/ferromagnet/oxide structures. *Nat. Nanotechnol.* **11**, 878–884 (2016).

15. Zhang, W. *et al.* Giant facet-dependent spin-orbit torque and spin Hall conductivity in the triangular antiferromagnet $IrMn_3$. *Sci. Adv.* **2**, (2016).

16. Chen, J.-Y. *et al.* Field-free spin-orbit torque switching of composite perpendicular CoFeB/Gd/CoFeB layers utilized for three-terminal magnetic tunnel junctions. *Appl. Phys. Lett.* **111**, 012402 (2017).

17. Nan, T. *et al.* Controlling spin current polarization through non-collinear antiferromagnetism. *Nat. Commun.* **11**, (2020).

18. Macneill, D. *et al.* Control of spin–orbit torques through crystal symmetry in WTe$_2$/ferromagnet bilayers. *Nat. Phys.* **13**, 300–306 (2017).

19. Shi, S. *et al.* All-electric magnetization switching and Dzyaloshinskii–Moriya interaction in WTe2/ferromagnet heterostructures. *Nat. Nanotechnol.* **14**, 945–949 (2019).

20. Song, P. *et al.* Coexistence of large conventional and planar spin Hall effect with long spin diffusion length in a low-symmetry semimetal at room temperature. *Nat. Mater.* **19**, 292–298 (2020).

21. Qi, X.-L. & Zhang, S.-C. Topological insulators and superconductors. *Rev. Mod. Phys.* **83**, 1057–1106 (2011).

22. Zhao, B. *et al.* Unconventional charge–spin conversion in weyl-semimetal WTe2. *Adv. Mater.* **2000818** (2020).

23. Baek, S. C. *et al.* Spin currents and spin–orbit torques in ferromagnetic trilayers. *Nat. Mater.* **17**, 509–513 (2018).

24. Lee, D. & Lee, K. Spin-orbit Torque Switching of Perpendicular Magnetization in Ferromagnetic Trilayers. *Sci. Rep.* **10**, 1772 (2020).

25. Li, X. *et al.* Materials Requirements of High-Speed and Low-Power Spin-Orbit-Torque Magnetic Random-Access Memory. *IEEE J. Electron Devices Soc.* **PP**, 1–1 (2020).

26. Mangin, S. *et al.* Current-induced magnetization reversal in nanopillars with perpendicular anisotropy. *Nat. Mater.* **5**, 210–215 (2006).

27. Han, J. *et al.* Room-temperature spin-orbit torque switching induced by a topological

insulator. *Phys. Rev. Lett.* **119**, 077702 (2017).

28. Li, P. *et al.* Magnetization switching using topological surface states. *Sci. Adv.* **5**, (2019).

29. Zhao, Z., Smith, A. K., Jamali, M. & Wang, J. P. External-field-free spin Hall switching of perpendicular magnetic nanopillar with a dipole-coupled composite structure. *Adv. Electron. Mater.* **6**, 1901368 (2020).

30. Krizakova, V., Garello, K., Grimaldi, E., Kar, G. S. & Gambardella, P. Field-free switching of magnetic tunnel junctions driven by spin–orbit torques at sub-ns timescales. *Appl. Phys. Lett.* **116**, 232406 (2020).

31. Wang, M. *et al.* Field-free switching of a perpendicular magnetic tunnel junction through the interplay of spin-orbit and spin-transfer torques. *Nat. Electron.* **1**, 582–588 (2018).

32. Coldea, M. *et al.* X-ray photoelectron spectroscopy and magnetism of Mn-Pd alloys. *J. Alloys Compd.* **417**, 7–12 (2006).

33. Avci, C. O. *et al.* Interplay of spin-orbit torque and thermoelectric effects in ferromagnet/normal-metal bilayers. *Phys. Rev. B* **90**, 224427 (2014).

34. Chen, Y.-T. *et al.* Theory of spin Hall magnetoresistance. *Phys. Rev. B* **87**, 144411 (2013).

35. Zhang, W., Han, W., Jiang, X., Yang, S.-H. & Parkin, S. S. P. Role of transparency of platinum–ferromagnet interfaces in determining the intrinsic magnitude of the spin Hall effect. *Nat. Phys.* **11**, 496–503 (2015).

36. Pai, C. F., Ou, Y., Vilela-Leão, L. H., Ralph, D. C. & Buhrman, R. A. Dependence of the efficiency of spin Hall torque on the transparency of Pt/ferromagnetic layer interfaces.

*Phys. Rev. B* **92**, 064426 (2015).

37. Zhang, W. *et al.* Spin Hall effects in metallic antiferromagnets. *Phys. Rev. Lett.* **113**, 196602 (2014).

38. Okada, A. *et al.* Spin-pumping-free determination of spin-orbit torque efficiency from spin-torque ferromagnetic resonance. *Phys. Rev. Appl.* **12**, 014040 (2019).

39. Zhu, L., Ralph, D. C. & Buhrman, R. A. Spin-Orbit Torques in Heavy-Metal-Ferromagnet Bilayers with Varying Strengths of Interfacial Spin-Orbit Coupling. *Phys. Rev. Lett.* **122**, 77201 (2019).

40. Feng, J. *et al.* Effects of oxidation of top and bottom interfaces on the electric, magnetic, and spin-orbit torque properties of Pt / Co / Al Ox trilayers. *Phys. Rev. Appl.* **13**, 044029 (2020).

41. Pai, C. F. *et al.* Spin transfer torque devices utilizing the giant spin Hall effect of tungsten. *Appl. Phys. Lett.* **101**, 122404 (2012).

42. Cho, S., Chris Baek, S., Lee, K.-D., Jo, Y. & Park, B.-G. Large spin Hall magnetoresistance and its correlation to the spin-orbit torque in W/CoFeB/MgO structures. *Sci. Rep.* **5**, 14668 (2015).

43. Zhu, L. *et al.* Strong damping-like spin-orbit torque and tunable Dzyaloshinskii–Moriya interaction generated byILow-resistivity $Pd_{1-x}Pt_x$ alloys. *Adv. Funct. Mater.* **29**, 1805822 (2019).

44. Tsai, H. *et al.* Electrical manipulation of a topological antiferromagnetic state. *Nature* **580**, 608 (2020).

45. Cao, J. *et al.* Spin-orbit torque induced magnetization switching in Ta/Co$_{20}$Fe$_{60}$B$_{20}$/MgO structures under small in-plane magnetic fields. *Appl. Phys. Lett.* **108**, 172404 (2016).

46. Baumgartner, M. *et al.* Spatially and time-resolved magnetization dynamics driven by spin-orbit torques. *Nat. Nanotechnol.* **12**, 980–986 (2017).

47. Avci, C. O. *et al.* Unidirectional spin Hall magnetoresistance in ferromagnet/normal metal bilayers. *Nat. Phys.* **11**, 570–575 (2015).

48. Lv, Y. *et al.* Large unidirectional spin Hall and Rashba-Edelstein magnetoresistance in topological insulator/magnetic insulator heterostructures. arXiv:1806.09066. (2018).

49. Guo, G. Y., Murakami, S., Chen, T.-W. & Nagaosa, N. Intrinsic spin Hall effect in platinum: first-principles calculations. **100**, 096401 (2008).

50. Sinova, J., Valenzuela, S. O., Wunderlich, J., Back, C. H. & Jungwirth, T. Spin Hall effects. *Rev. Mod. Phys.* **87**, 1213–1260 (2015).
**Acknowledgments**

This research was supported in part by ASCENT, one of six centers in JUMP, a Semiconductor Research Corporation (SRC) program sponsored by DARPA. The authors thank the NSF Center for Energy Efficient Electronics Science (E3S) and TSMC for financial support. Part of this work was performed at the Stanford Nano Shared Facilities (SNSF)/Stanford Nanofabrication Facility (SNF), supported by the National Science Foundation under award ECCS-1542152. The research at the University of Nebraska-Lincoln is supported by the National Science Foundation through the Nebraska Materials Science and Engineering Center (MRSEC, Grant No. DMR-1420645). P. Q. acknowledges support from the National Research Council Research Associateship Program. S.E. and M. B. V. acknowledge funding from NSF award DMR-

1905909, and assistance from Randy Dumas at Quantum Design with VSM measurements. The authors would also like to acknowledge Dr. Carlos H. Diaz, Dr. Peng Li, and Dr. Juliet Jamtgaard for fruitful discussions.

**Author contributions**

M. DC conceived, designed, and coordinated the research with contributions from M.M., S.-J.L., W.T., and S.X.W. S.X.W supervised the study. M. DC grew thin films, performed XRD measurement, fabricated Hall bar, ST-FMR device, carried out ST-FMR, SHH, and switching measurements with contributions from Y.D., X.L., C.B., F.X., and Y-L. H. D-F. S. and E. T. performed DFT calculations. V.D.H., A. H., and W. W. carried out TEM and EDS studies. A.V. performed pole figure measurements. M. B.V. and S.E. performed magnetometry measurements. P.Q., B.K., and J.B. performed PNR measurements and modelling. M. DC performed LLG simulations. M. DC performed data analysis and wrote the manuscript with contributions from D-F. S, P.Q., A.V., and S.X.W. All authors discussed the results and commented on the manuscript.

**Online content**

Any methods, additional references, Nature Research reporting summaries, source data, extended data, supplementary information, acknowledgements, peer review information; details of author contributions and competing interests; and statements of data and code availability are available online.

**Competing Interests**

The authors declare no competing interests.

**Data Availability**

The data that support the findings of this study are available from the corresponding authors on reasonable request.